\documentclass[a4paper]{amsart}

\usepackage[utf8]{inputenc}
\usepackage[T1]{fontenc}

\usepackage{siunitx}
\DeclareSIUnit\wtpercent{wt.\percent}

\usepackage{changepage}

\usepackage{booktabs}
\usepackage{threeparttable}

\usepackage{amssymb} 
\usepackage{amsmath} 
\DeclareMathOperator*{\mean}{mean} 

\usepackage{textcomp}
\usepackage{adjustbox}
\usepackage{graphicx}

\usepackage{placeins}

\usepackage{lineno}

\theoremstyle{definition}

\theoremstyle{remark}

\numberwithin{equation}{section}

\usepackage[foot]{amsaddr}

\begin{document}


\title[Magnetic Lenz lenses increase the LOD in NMR]{Magnetic Lenz lenses increase the limit-of-detection in nuclear magnetic resonance}


\author[N. Spengler]{Nils Spengler$^{1,2,\dagger}$}
\address{$^1$IMTEK – Department of Microsystems Engineering, University of Freiburg, 79110 Freiburg, Germany}
\address{$^2$Institute of Microstructure Technology (IMT), Karlsruhe Institute of Technology, 76344 Eggenstein-Leopoldshafen, Germany}
\thanks{$^\dagger$N.S. and P.T.W contributed equally to this work.}

\author[P. T. While]{Peter T. While$^{1,3,\dagger}$}
\address{$^3$Department of Radiology and Nuclear Medicine, St. Olav's University Hospital, 7030 Trondheim, Norway}

\author[M. V. Meissner]{Markus V. Meissner$^{1,2}$}

\author[U. Wallrabe]{Ulrike Wallrabe$^1$}

\author[J. G. Korvink*]{Jan G. Korvink$^{1,2,*}$}
\email{jan.korvink@kit.edu}


\keywords{Lenz lens $|$ magnetic flux focusing $|$ Faraday induction $|$ sensitivity $|$ MR microscopy}


\dedicatory{}

 \begin{abstract}
 A high NMR detection sensitivity is indispensable when dealing with mass and volume-limited samples, or whenever a high spatial resolution is required. The use of miniaturised RF coils is a proven way to increase sensitivity, but may be impractical and is not applicable to every experimental situation. We present the use of magnetic lenses, denoted as Lenz lenses due to their working principle, to focus the magnetic flux of a macroscopic RF coil into a smaller volume and thereby locally enhance the sensitivity of the NMR experiment -- at the expense of the total sensitive volume. Besides focusing, such lenses facilitate re-guiding or re-shaping of magnetic fields much like optical lenses do with light beams. For the first time we experimentally demonstrate the use of Lenz lenses in magnetic resonance and provide a compact mathematical description of the working principle. Through simulations we show that optimal arrangements can be found.
 \end{abstract}

\maketitle

Both nuclear magnetic resonance (NMR) spectroscopy and magnetic resonance imaging (MRI) suffer from an inherently low sensitivity. The signal strength is primarily determined by the equilibrium Boltzmann distribution, with energy levels of the spin states just slightly above the thermal energy. Consequently, the limit of detection (LOD) is up to ten orders of magnitude worse compared to other analytical techniques \cite{webb_radiofrequency_1997,lacey_high-resolution_1999}. This fact severely limits the lowest detectable quantity in NMR spectroscopy and the highest achievable spatial resolution in MRI, both being directly proportional to the signal-to-noise ratio (SNR) of the experiment. The SNR, first derived by Hoult and Richards \cite{hoult_signal--noise_1976}, is given by \cite{peck_design_1995}
\begin{equation}
\label{eqn:snr}
\mathrm{SNR}=\frac{k_0\frac{B_1}{i}V_{\mathrm{obs}}N\gamma\hbar^2I\left(I+1\right)\frac{\omega_0^2}{k_\mathrm{B}T3\sqrt{2}}}{\sqrt{4k_\mathrm{B}TR\Delta{}f}}\propto\frac{B_1}{i\sqrt{R}}=\frac{B_1}{\sqrt{P}},
\end{equation}
where $k_0$ is a scaling factor to account for the homogeneity of the radio-frequency (RF) coil employed, $B_1$ is the RF-coil's magnetic field strength, $i$ the unit current, $B_1/i$ the coil's sensitivity, $V_{\mathrm{obs}}$ the observed sample volume, $k_\mathrm{B}$ is the Boltzmann constant, $T$ is the temperature, $N$ the spin density, $I$ the spin quantum number, $\hbar$ is Planck's constant $h$ divided by $2\pi$, $\omega_0=-\gamma{}B_0$ is the Larmor frequency determined by the gyromagnetic ratio $\gamma$ of the nucleus of interest and the strength of the static magnetic field $B_0$, $R$ is the electrical resistance of the coil contributing thermal noise, $\Delta{}f$ is the bandwidth of the receiver, and $P$ the power.

The search for higher SNR has resulted in steadily increasing $B_0$-field strengths, currently reaching a maximum of around \SI{23.5}{\tesla} (\SI{1}{\giga\hertz} $^1$H Larmor frequency) for commercially available NMR systems, while field strengths of human scale MRI magnets typically do not exceed \SI{3}{\tesla} in the clinic.
While both the noise level and the Boltzmann distribution would benefit from a reduced sample temperature, the freezing point and the operating temperature of the sample impose practical limits. Non-equilibrium Boltzmann polarisation factors can also be reached through spin 
order transfer techniques such as dynamic nuclear polarisation (DNP) or parahydrogen induced polarisation (PHIP), which, however, often involve toxic substances and hence are not yet generally compatible with living biological samples.

Consequently, the SNR is maximised by using optimised hardware along the receive path, such as dedicated RF receiver coils. Such coil designs follow two strategies: (i)~the filling factor \cite{hill_limits_1968} and hence the magnetic interaction is maximised when the coil geometrically conforms to the sample as closely as possible, and (ii)~the sensitivity increases linearly as the diameter of the coil decreases for a constant height-to-diameter ratio \cite{peck_design_1995}. The relative intrinsic SNR is therefore higher when using a smaller, sample-adapted coil. Hence such coils are used when recording MR images of body parts, e.g., the brain, or when acquiring spectra from rare or volume-limited substances \cite{olson_high-resolution_1995}. However, additional equipment is required to cover the various cases, which is cost-intensive and requires maintenance, while situations remain where it is not yet possible to place the sample inside a dedicated active coil, such as when studying inner parts like organs.

\begin{figure}[h!tb]
\begin{adjustwidth}{-2.5cm}{0in}
\centering
\includegraphics[]{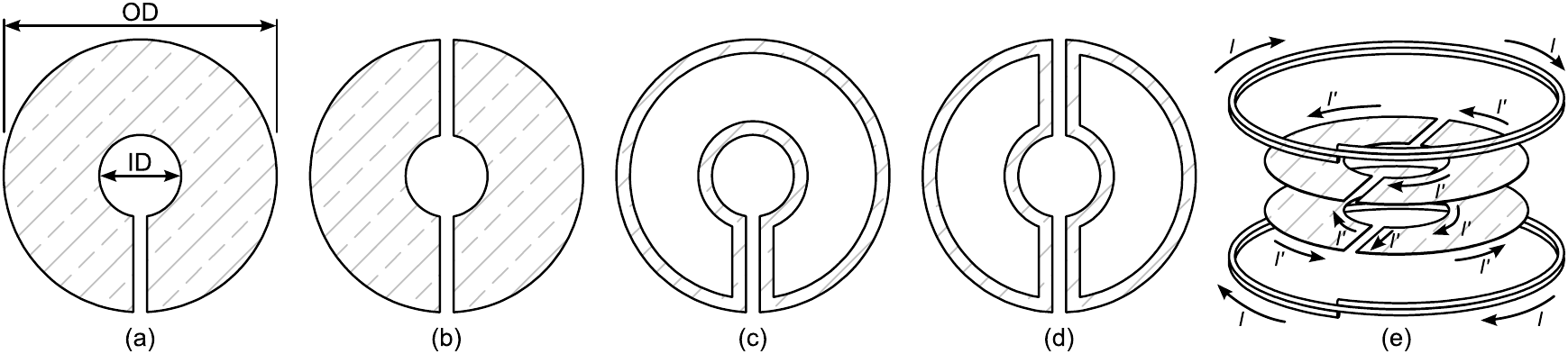}
\caption{The four basic designs of Lenz lenses, made from either solid metal (subfigures (a) and (b)) or wire material (sub-figures (c) and (d)), arranged symmetrically ((b) and (d)) or non-symmetrically ((a) and (c)). The slit(s) guide induced current $I'$ from the outer edge to the inner edge, while reversing the flow direction, as further depicted in (e), where two lenses are arranged in parallel in a Helmholtz pair like configuration, denoted as a double Lenz lens configuration.}
  \label{fig:ll-overview}
\end{adjustwidth}
\end{figure}

To circumvent these issues, we introduce here a method to locally improve the sensitivity of the NMR (or MRI) experiment by means of broadband passive magnetic lenses. The working principle behind these lenses is geometry based and governed by Lenz' law, and hence they are referred to as Lenz lenses \cite{schoenmaker_magnetic_2013}. Such lenses are capable of focusing the magnetic field of a macroscopic RF coil into a smaller spatial region, thereby locally increasing the flux density for the same applied radiofrequency current. Lenz lenses can be simply made from wire, or conductive sheets such as copper foil, as shown in Fig.~\ref{fig:ll-overview}, which illustrates two basic shapes.

Due to their simple geometry and wireless, broadband inductive coupling, Lenz lenses can easily be tailored to the intended application in a much more flexible manner than RF resonators, and are straightforward to fabricate. Multiple lenses can be combined to enable the shaping or rotating of the magnetic field, similar to lenses in optics, and hence introduce a whole new degree of flexibility for the MR analysis of diverse samples.

\FloatBarrier
\section{Results}
\subsection{Theory}
\label{sec:theo}
Expressions are now derived for calculating the induced current within an arbitrary number of circular Lenz lenses made from wires and placed within the incident magnetic field generated by an external RF coil. Note that, by the theory of reciprocity \cite{HOULT:2000fd,HOULT:2011jv}, the expressions may be used in an equivalent sense to describe a situation in which a distinct volume of precessing magnetisation induces current within the Lenz lenses, which in turn induce current within the external RF coil. That is, the Lenz lenses and corresponding theory are applicable to both the excitation and receive chains of a conventional NMR (or MRI) experiment.

Consider the simple arrangement of placing a single circular Lenz lens coaxially at the midpoint between two elements of a Helmholtz pair. Faraday's law of induction states that the electromotive force (emf), $\epsilon$, generated in the lens is equal to the negative rate of change of magnetic flux $\Phi$ that it encloses:
\begin{equation}
\epsilon=-\frac{\partial\Phi}{\partial t}.\label{Faraday}
\end{equation}
Lenz's law states that the sense of the current induced is such that it opposes this flux, hence the minus sign in equation~(\ref{Faraday}). In the limit that the gap size between the elements connecting the outer and inner loops of the lens is zero, the flux impressed on the lens is equal to that enclosed by the outer loop minus that enclosed by the inner loop:
\begin{equation}
\Phi=2\left(M_{BH}-M_{SH}\right)I_{H},\label{PhiB}
\end{equation}
where $I_{H}$ is the current in the Helmholtz pair, $M_{BH}$ is the mutual inductance between the outer loop of the lens ($B\equiv\textrm{big}$) and one element of the Helmholtz pair, and a similar definition holds for $M_{SH}$ ($S\equiv\textrm{small}$). We ignore capacitive effects and model the Lenz lens itself as an $RL$-circuit:
\begin{equation}
V=I_{L}R_{L}+L_{L}\frac{dI_{L}}{dt},\label{Voltage}
\end{equation}
where $L_{L}=L_{S}+L_{B}-2M_{SB}$ is the self-inductance of the lens (with corresponding definitions for $L_{S}$, $L_{B}$ and $M_{SB}$), $R_{L}$ is the resistance of the lens and $I_{L}$ is the induced current. Equating the potential in equation~(\ref{Voltage}) to the emf in equation~(\ref{Faraday}) (combined with equation~(\ref{PhiB})) and assuming a time-harmonic regime, we obtain the following expression for the induced current within the lens:
\begin{equation}
I_{L}=\frac{2i\omega\left(M_{BH}-M_{SH}\right)I_{H}}{\left[R_{L}-i\omega\left(L_{S}+L_{B}-2M_{SB}\right)\right]},\label{IL1}
\end{equation}
where $\omega$ is the operating frequency (radians/s) of the RF coil. Equation~(\ref{IL1}) represents a generalisation of an expression provided by Schoenmaker et al. \cite{schoenmaker_magnetic_2013}, which was derived for a Lenz lens operating in the kilohertz regime. Those authors ignored the resistance term in equation~(\ref{IL1}), hence neglecting the frequency dependence of the induced current, and they also ignored the mutual inductance between the inner and outer loops of the lens.

For high-frequency applications, such as the megahertz regime relevant to NMR, it is necessary to consider the skin-effect in the resistance and inductance calculations used in equation~(\ref{IL1}). Let us label the resistance of a wire under direct current to be $R_{0}$ and under (high-frequency) alternating current to be $R_{L}$, as above. The ratio $R_{L}/R_{0}$ can be calculated exactly using Kelvin-Bessel functions \cite[p.185]{RamoWV1994}, however this approach is computationally unstable when the skin-depth is small relative to the wire radius. A common alternative is to approximate the relative increase in resistance by the inverse of the relative decrease in the effective cross-sectional area of the ring defined by the skin-depth \cite{Wheeler1942PIRE}. This approach is accurate to within 5.5\% of the exact result when the ratio of wire radius, $a$, to skin-depth, $\delta$, is greater than one \cite{Knight2013}, and below this limit the resistance is simply equal to $R_{0}$:
\begin{eqnarray}
\frac{R_{L}}{R_{0}}=\left\{\begin{array}{cc}1,&0\leq\frac{a}{\delta}<1\\
\frac{a^{2}}{2a\delta-\delta^{2}},&\frac{a}{\delta}\geq1,
\end{array}\right.\label{Rac}
\end{eqnarray}
where $R_{0}=\rho\left[2\pi r_{B}+2\pi r_{S}+2(r_{B}-r_{S})\right]/(\pi a^{2})$, in which $r_{B}$ is the radius of the outer loop of the lens and $r_{S}$ is the radius of the inner loop, $\delta=\sqrt{2\rho/\omega\mu}$, and $\rho$ and $\mu$ are the resistivity and permeability of the conductor, respectively (e.g. for copper $\rho=1.68\times10^{-8}$ $\Omega$m and $\mu\approx\mu_{0}=4\pi\times10^{-7}$ H/m).

The self-inductance terms, $L_{B}$ and $L_{S}$, may be calculated using the approximation \cite{Dengler2013}:
\begin{equation}
L\approx\mu_{0}r\left[\ln\left(\frac{8r}{a}\right)-2+\frac{Y}{2}\right],\label{L}
\end{equation}
where $r$ is the radius of the loop in question and the parameter $Y$ depends on the frequency of operation. $Y=0$ is the high-frequency limit for which current is restricted to the surface of the wire and $Y=1/2$ is the low-frequency limit for which the current is uniform throughout the wire; hence we set $Y=R_{0}/(2R_{L})$ (see equation~(\ref{Rac})). Note that equation~(\ref{L}) is accurate to $O(a^{2}/r^{2})$ \cite{Dengler2013}.

The mutual inductances between the Helmholtz coil and the elements of the Lenz lens, and between the lens elements themselves, can be calculated using the formulae provided by Babic et al. \cite{BabicSAG2010IEEETM} (i.e. equations~(24)-(25)). These formulae were derived following a vector potential argument and are applicable to any pair of arbitrarily placed circular conductors. Note that Matlab code for evaluating these formulae has been made available by Babic et al. on their publisher's website (http://ieeexplore.ieee.org).

Let us now consider the general case of multiple lenses placed arbitrarily within the vicinity of a Helmholtz pair, as depicted in Fig.~\ref{fig:ll-overview}(e). We must now treat the two loops of the Helmholtz pair separately (subscripts $H1$ and $H2$) and consider also the mutual inductances that exist between the elements of different lenses. After careful consideration of the enforced current sense between inner and outer loops of each lens, we arrive at the following expression for the flux contained by the $k$th lens ($k=1:K$):
\begin{equation}
\begin{array}{ll}
 \Phi_{k}&=\left(M_{BkH1}+M_{BkH2}-M_{SkH1}-M_{SkH2}\right)I_{H}\\
&+\sum^{K}_{j=1,j\neq k}\left(M_{BkBj}-M_{SkBj}-M_{BkSj}+M_{SkSj}\right)I_{Lj},
\end{array}
\label{Phik}
\end{equation}
where, for example, $M_{BkSj}$ is the mutual inductance between the outer loop of the $k$th lens and the inner loop of the $j$th lens, and $I_{Lj}$ is the induced current in the $j$th lens. Similarly, the potential for the $k$th lens is given by:
\begin{equation}
V_{k}=\left[R_{Lk}-i\omega\left(L_{Sk}+L_{Bk}-2M_{SkBk}\right)\right]I_{Lk}.\label{Vk}
\end{equation}
Combining equations(\ref{Faraday}),~(\ref{Phik}) and~(\ref{Vk}) and rearranging, we obtain the following matrix equation:
\begin{equation}
\mathcal{L}\textbf{\textit{I}}_{L}=\textbf{\textit{H}},\label{ILm}
\end{equation}
where
\begin{eqnarray}
\begin{array}{lll}
\mathcal{L}_{kk}&=&R_{Lk}-i\omega\left(L_{Sk}+L_{Bk}\right.\\&&\left.-2M_{SkBk}\right)\\
\mathcal{L}_{kj}&=&-i\omega\left(M_{BkBj}-M_{SkBj}\right.\\&&\left.-M_{BkSj}+M_{SkSj}\right)\\
H_{k}&=&i\omega\left(M_{BkH1}+M_{BkH2}\right.\\&&\left.-M_{SkH1}-M_{SkH2}\right)I_{H}\end{array}\quad\left(\!\!\!\begin{array}{c}k=1:K\\
j=1:K\\
j\neq k\end{array}\!\!\!\right),\label{ILme}
\end{eqnarray}
and $\textbf{\textit{I}}_{L}$ is a vector of length $K$, which contains the induced current for each lens. Note that it is straightforward to show that equations~(\ref{ILm})-(\ref{ILme}) reduce to equation~(\ref{IL1}) for the case of a single symmetrically placed lens.

\subsection{Simulations and optimisation}\label{ssec:meth_simopt}
Equations~(\ref{ILm})-(\ref{ILme}) permit the investigation of the dependence of the induced current(s) and corresponding magnetic field on a variety of design parameters, such as the number of lenses, their geometry and the frequency of operation. As an illustrative example, equation~(\ref{ILm}) was solved for several different arrangements appropriate to the experiments described in Section \ref{sec:expts}. That is, we considered a Helmholtz pair with a radius of 0.6 mm and a separation of 0.6 mm, carrying a current of 66 mA at 500 MHz. The wire for the lens(es) was modelled using an effective circular cross-section with radius $a=8.9$ $\mu$m. The radius of the outer loop of the lens(es) was fixed at $r_{B}=0.5$ mm for all cases. The radius of the inner loop of the lens(es), $r_{S}$, was set at either 0.1 mm or 0.2 mm and we considered both single lens and double lens cases, with a lens separation for the latter of either 0.1 mm or 0.2 mm. All simulations were performed using Matlab\textsuperscript{\textregistered} (R2012b, Mathworks\textsuperscript{\textregistered}).

In general, a smaller radius for the inner loop of a lens results in a stronger magnetic field at the midpoint. However, in NMR (or MRI) it is desirable not only to have a strong transmit RF field (and correspondingly high receive sensitivity) but also to have a homogeneous field, such that the flip angle is constant throughout the region of interest (ROI). Therefore the design of the Lenz lens(es) for NMR becomes an optimisation problem with the goal of maximising the induced field within the ROI, with respect to the lens geometry, while maintaining an acceptable level of homogeneity. That is, we wish to solve:
\begin{eqnarray}
&\max_{\boldsymbol{\rho}}\left\{\mean_{\textrm{ROI}}\left\{B_{z}(\boldsymbol{\rho})\right\}\right\}\quad\mathrm{s.t.}\nonumber\\&\quad\mean_{\textrm{ROI}}\left\{\left\|B_{z}(\boldsymbol{\rho})-\mean_{\textrm{ROI}}\left\{B_{z}(\boldsymbol{\rho})\right\}\right\|\right\}\leq\epsilon,\label{opt}
\end{eqnarray}
where $\boldsymbol{\rho}$ is a vector containing the radii of the inner and outer loops and the axial positions of each lens, $B_{z}$ is the axial component of the total magnetic field calculated using the Biot Savart law, and $\epsilon$ is the specified acceptable average field error. As an illustrative example in keeping with the simulations outlined above, the ROI was chosen to be a cylindrical volume of radius 0.06 mm and height 0.06 mm, centred at the origin and coaxial with the lenses and Helmholtz coil. The mean terms in equation~(\ref{opt}) were evaluated by integrating numerically (trapezoidal rule) over the ROI and then dividing by the volume.

The optimisation in equation~(\ref{opt}) was achieved using the function \texttt{fmincon} from the Matlab\textsuperscript{\textregistered} Optimisation Toolbox\texttrademark\thinspace (interior-point algorithm) for three cases: one, two and four lenses. To improve convergence for the cases with multiple lenses, the optimisation was performed in two stages: firstly, the axial placements of the lenses were fixed at equal increments between the Helmholtz coils and equation~(\ref{opt}) was solved with respect to the lens radii alone; secondly, the results from the first optimisation were used as the starting guess for the full optimisation of equation~(\ref{opt}). Additional constraints were imposed to ensure that: the inner loops were always smaller than the outer loops; the axial placements of multiple lenses were always separated by at least the wire diameter; the axial placements of pairs of multiple lenses were always symmetric about $z=0$ and closer to the origin than the Helmholtz coils; the radii of the outer loops were always less than twice that of the Helmholtz coil (and all radii were positive). The optimisations took the order of seconds (1 lens, 7 s; 2 lenses, 25 s; 4 lenses, 106 s) on a standard desktop computer (2.7 GHz Intel\textsuperscript{\textregistered} Core\texttrademark\thinspace  i7 processor with 16 GB of RAM).

\begin{figure}[h!tbp]
\centering
\includegraphics[]{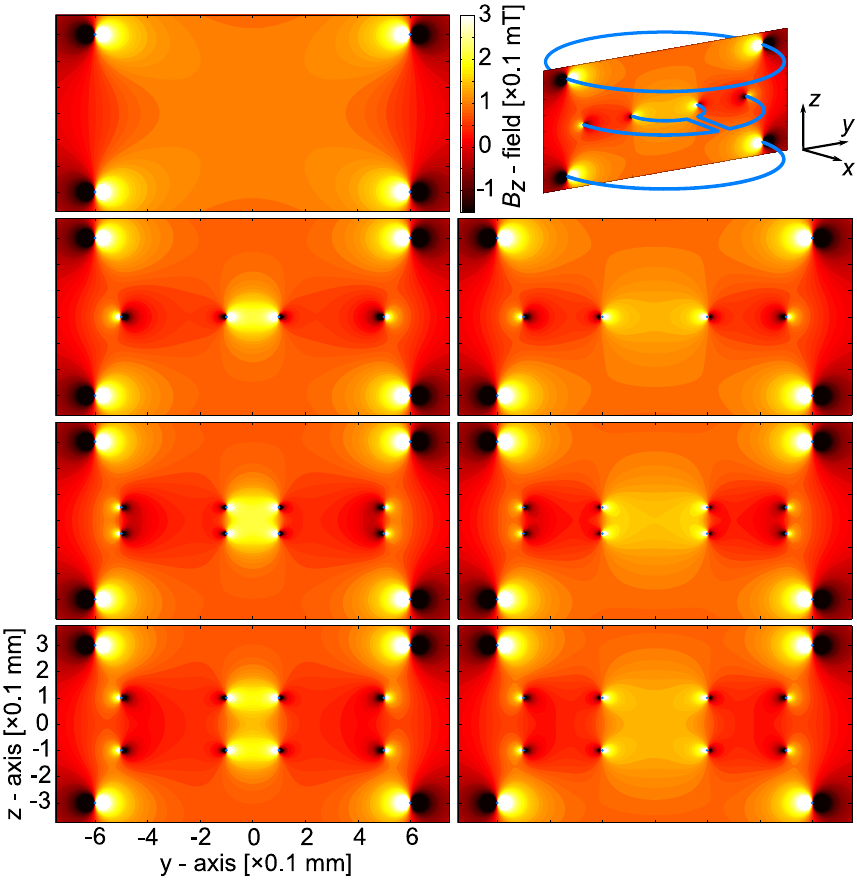}
 \caption{The $B_{z}$-field induced by a Helmholtz pair and a variety of different lenses: i.e. zero, one or two lenses of different inner loop size and/or axial position. The colour contour plots are all plotted on the $yz$-plane as depicted in the illustrative example at the upper right of the figure.}
\label{simfigs}
\end{figure}

\begin{figure}[h!tbp]
\centering
\includegraphics[]{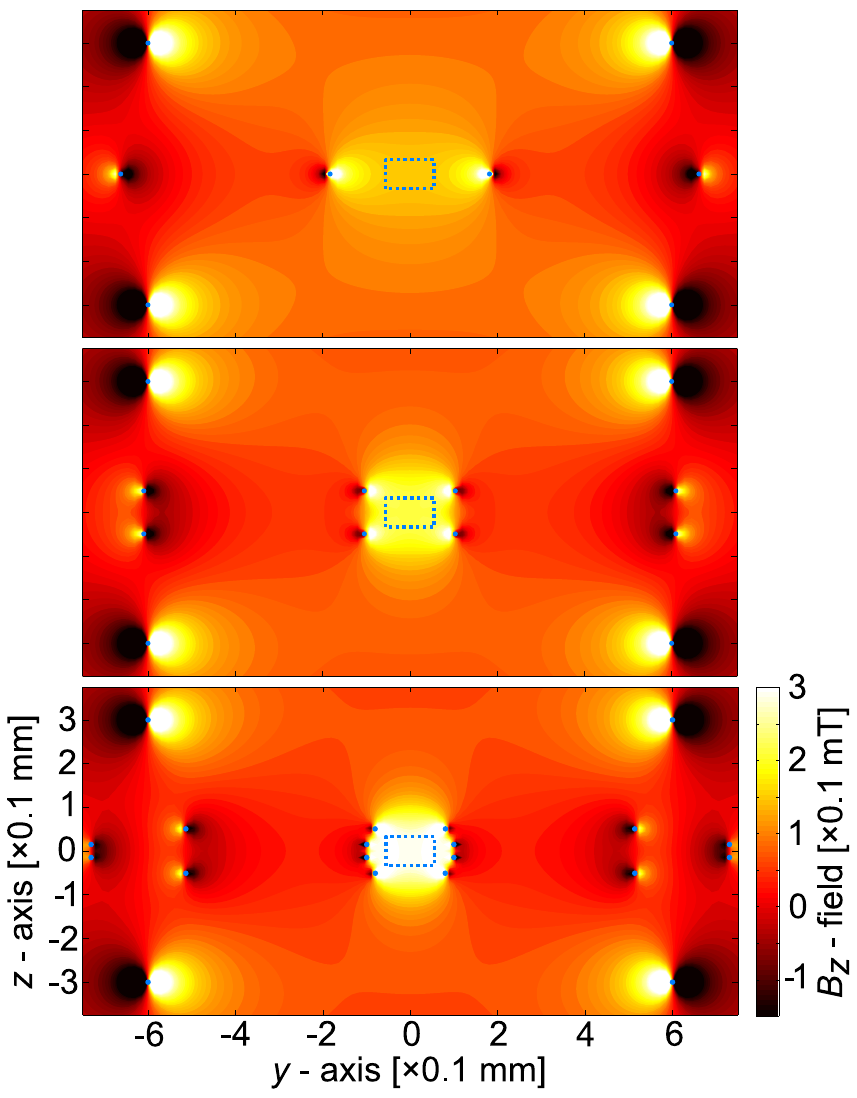}
 \caption{Three optimised systems consisting of a Helmholtz pair and either one, two or four lenses. The size and position of the lens elements have been optimised to maximise the amplification within a cylindrical volume, depicted in the $yz$-plane by the blue dashed rectangle, whilst constraining the average field error to be 1\% within this region of interest. The example with four lenses exhibits a threefold amplification of field sensitivity.}
\label{optfigs}
\end{figure}

Fig.~\ref{simfigs} displays colour contour plots of the simulated $z$-component of the induced magnetic field over the plane $x=0$ for several different arrangements of Lenz lenses within the Helmholtz pair. Similarly, Fig.~\ref{optfigs} displays the results of solving the optimisation problem defined by equation~(\ref{opt}) for the cases of one, two and four lenses.

\FloatBarrier
\subsection{MRI experiments}\label{sec:expts}
Symmetrical wire and plate-based Lenz lenses according to Figs.~\ref{fig:ll-overview}(b) and \ref{fig:ll-overview}(d) with \SI{1.0}{\mm} outer diameter (OD), and inner diameters (IDs) of \SIlist{0.2;0.4}{\mm} were patterned by means of photo-lithography and electroplating on glass substrates to be used with a custom micro Helmholtz coil pair. The four designs fabricated are summarised in Table~\ref{tab:table1}, and short identifiers were assigned for easier discrimination of the different designs.
\begin{table}[h!tb]
\centering
\caption{Overview of the four different double Lenz lens variants manufactured.}
\begin{tabular}{llrrr}
Type & Reference & OD (mm) & ID (mm) & Identifier \\
\midrule
Plate & Fig.~\ref{fig:ll-overview}(b) & 1.0 & 0.2 & LL1  \\
 & & & 0.4 & LL2  \\
\cmidrule{1-5}
Wire & Fig.~\ref{fig:ll-overview}(d) & 1.0 & 0.2  & LL3  \\
 & & & 0.4 & LL4  \\
\bottomrule
\end{tabular}
\label{tab:table1}
\end{table}

To quantify the amplification of the lenses and to verify the theory developed above, we acquired a series of spin echo imaging experiments using a similar \SI{1.2}{\mm} diameter micro Helmholtz coil pair setup as presented in \cite{spengler_heteronuclear_2016}, but with one instead of two coil windings on each side. The Lenz lens designs LL1-LL4 were primed with DI-H$_2$O and arranged in the setup illustrated in Fig.~\ref{fig:ll-overview}(e) before MRI was performed at steadily decreasing attenuations. For reference, we further acquired MR images without a Lenz lens present using comparable chips filled with H$_2$O. A photograph of the actual setup and a brief summary of the results obtained is presented in Supplementary Fig.~1.

Pulse parameters for the individual designs were determined by exporting the absolute signal values from ParaVision using a macro (Bruker).
Keeping the pulse length $\tau$ constant, the $B_1$-field required to generate a flip angle $\alpha$ for a specific nucleus is given by
\begin{equation}
B_1=\frac{\alpha}{\gamma\tau},\label{b1}
\end{equation}
which leads to $B_1=\SI{0.53}{\milli\tesla}$ for $\tau=\SI{11}{\micro\second}$, and $\alpha=\pi/2$ (\SI{90}{\degree}) using the proton gyromagnetic ratio $\gamma{}(^1\mathrm{H})=\SI{267.513e6}{\radian\per\second\per\tesla}$.

However, due to the non-uniform field profile of the lens, a mixture of flip angles is generated. A global \SI{90}{\degree}-pulse was defined by locating the highest signal amplitude along the sweeped power range, although the local flip angle might be \SI{\leq 90}{\degree} at these particular pulse parameters.

The probe efficiency $\eta_{\mathrm{p}}$, i.e., the conversion efficiency of power into magnetic field for a certain probe (Lenz lens) is given by
\begin{equation}
\eta_{\mathrm{p}}=\frac{B_1}{\sqrt{P}}\label{etap}
\end{equation}
and therefore $\eta_{\mathrm{p}}\propto\mathrm{SNR}$ according to equation~(\ref{eqn:snr}).

\begin{figure}[h!tbp]
\centering
\includegraphics[]{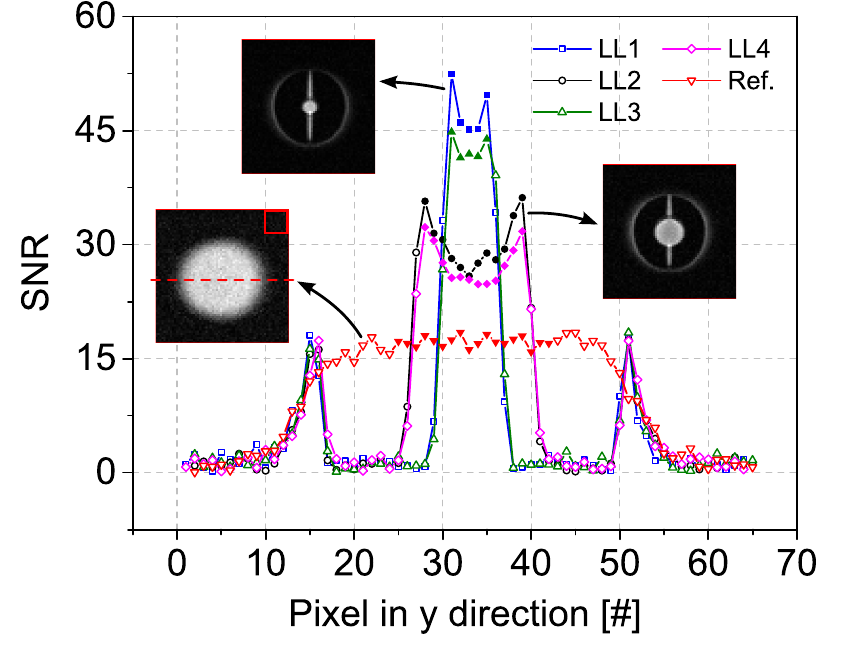}
\caption{Extracted SNR line profiles of LL1-LL4 and of the reference scan along the $y$-axis, as exemplified by the broken line in the reference image. All scans were acquired at a global flip angle of \SI{90}{\degree}. The mean noise was calculated from a \num{12 x 12} pixel matrix in the upper right corner of the same scan, as depicted. Statistics were derived from the regions illustrated by filled symbols. The powers for LL1 to LL4 and the reference scan were \SIlist{0.18;0.25;0.18;0.32;1}{\watt}.}
  \label{fig:90deg}
\end{figure}

Fig.~\ref{fig:90deg} presents SNR line profiles along the horizontal $y$-axis for LL1-LL4 and the reference scan without a lens. The SNR was calculated by taking the ratios of the exported absolute values and the mean noise value obtained from a \num{12 x 12} pixel matrix from the reference scan, as indicated in the figure. The line profiles were extracted from the images that were acquired using the individual, global \SI{90}{\degree}-pulse parameters evaluated for each type. For LL1-LL4 the determined powers $P_i$ were \SIlist{0.18;0.25;0.18;0.32}{\watt}, while for the reference scan, the \SI{90}{\degree}-pulse was found at $P_0=\SI{1}{\watt}$. These values lead to probe efficiencies $\eta_{\mathrm{p},i}$ of \SIlist{1.25;1.06;1.25;0.94}{\milli\tesla\,\watt^{-1/2}}, for LL1-LL4 respectively, while the reference efficiency was $\eta_{p,0}=\SI{0.53}{\milli\tesla\,\watt^{-1/2}}$.

\begin{figure}[h!tbp]
\centering
\includegraphics[]{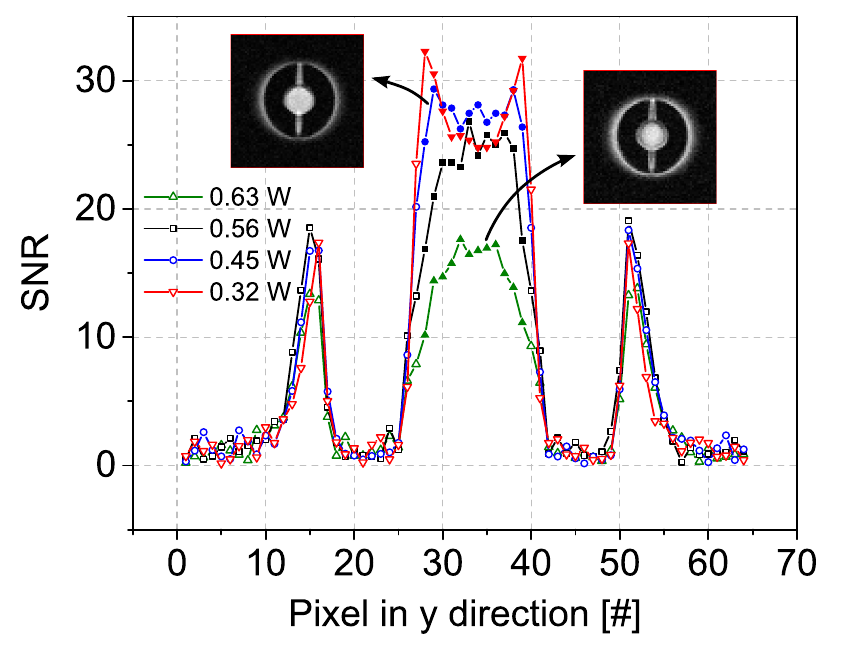}
    \caption{SNR profiles of LL4 at global flip angles of \SIlist{90;106;119;126}{\degree} and corresponding pulse powers of \SIlist{0.32;0.45;0.56;0.63}{\watt}. Statistics were derived from the regions illustrated by filled symbols.}
  \label{fig:LL4}
\end{figure}

Note that the non-uniform field profile associated with each lens leads to a local sinusoidal modulation of the flip angle while sweeping the power. As such, while the global \SI{90}{\degree}-pulse defined above results in the highest achievable signal amplitude, it does not lead to the highest achievable signal uniformity within the ROI. The evolution of the SNR profile of LL4 for $\alpha=\text{\SIlist{90;106;119;126}{\degree}}$ is illustrated in Fig.~\ref{fig:LL4}, where $\alpha=\SI{90}{\degree}$ was found at \SI{0.32}{\watt}, and the profile clearly changes from convex to concave shape as the power is increased.

The experimental results shown in Figs.~\ref{fig:90deg} and \ref{fig:LL4} are summarised in Table~\ref{tab:table2}, which furthermore contains calculated figures of merit. Data points defining the regions of interest, which were taken into account for calculations, are marked with filled symbols in both figures, while those values outside the ROIs are represented by hollow symbols.

\begin{table}[h!tbp]
\begin{adjustwidth}{-1.69cm}{0in}
\centering
\begin{threeparttable}
\caption{Calculated mean, maximum, and standard deviation (STD) of the SNR values from Fig.~\ref{fig:90deg} and Fig.~\ref{fig:LL4} as well as derived figures of merit.}
\label{tab:table2}
\begin{tabular}
{@{\extracolsep{\fill}}lrrrrrrrrrrrr}
Design & $i$\tnote{$*$} & $P_i$ (\si{\watt}) & $\alpha$\tnote{$\dagger$} & Mean & Max        & STD       & $n_i$\tnote{$\ddagger$} & $\sqrt{P_0/P_i}$ & $M_i$\tnote{$\mathsection$} & $\eta_{\mathrm{p},i}$ $\left(\frac{\si{\milli\tesla}}{\sqrt{\si{\watt}}}\right)$  & $\eta_{\mathrm{eff},i}$ $\left(\frac{\si{\milli\tesla}}{\sqrt{\si{\watt}}}\right)$\tnote{$\mathparagraph$} & ${V_i}/{V_0}$\tnote{$\#$}\\
\midrule
Ref.  & 0 & {1.00} & \SI{90}{\degree}  & {17.2} & {18.4} & {0.7} & 18 & {1.0} & {1.0} & {0.53} & {0.53} & {1}\\
LL1 & 1 & {0.18} & \SI{90}{\degree}  & {47.7} & {52.4} & {3.2} &  5 & {2.4} & {2.8} & {1.25} & {1.48} & {1/36}\\
LL2 & 2 & {0.25} & \SI{90}{\degree}  & {30.2} & {36.2} & {3.4} & 12 & {2.0} & {1.8} & {1.06} & {0.95} & {1/9}\\
LL3 & 3 & {0.18} & \SI{90}{\degree}  & {42.7} & {44.8} & {1.5} &  5 & {2.4} & {2.5} & {1.25} & {1.33} & {1/36}\\
LL4 & 4 & {0.32} & \SI{90}{\degree}  & {27.5} & {32.3} & {2.8} & 12 & {1.8} & {1.6} & {0.94} & {0.85} & {1/9}\\
\cmidrule{1-13}
LL4 & 5 & {0.45} & \SI{106}{\degree} & {24.5} & {29.3} & {1.2} & 12 & {1.5} & {1.4} & {0.94} & {0.74} & {1/9}\\
LL4 & 6 & {0.56} & \SI{119}{\degree} & {23.2} & {26.8} & {3.2} & 12 & {1.3} & {1.3} & {0.94} & {0.69} & {1/9}\\
LL4 & 7 & {0.63} & \SI{126}{\degree} & {15.0} & {17.6} & {2.4} & 12 & {1.3} & {0.9} & {0.94} & {0.48} & {1/9}\\
\bottomrule
\end{tabular}
\begin{tablenotes}
\item[$*$] Running index number.
\item[$\dagger$] A global $\alpha$ of \SI{90}{\degree} was assumed at the maximum peak SNR amplitude. Deviating flip angles were calculated based on $\eta_{\mathrm{p},i}$ and by combining equations~(\ref{b1}) and (\ref{etap}).
\item[$\ddagger$] Number of voxels $n_i$ taken to calculate mean, max and STD. The regions taken into account are represented by filled symbols in Figs.~\ref{fig:90deg} and \ref{fig:LL4}.
\item[$\mathsection$] Enhancement $M_i$ based on the ratio of the mean SNR values with respect to the reference scan without any lens.
\item[$\mathparagraph$] Effective probe efficiency $\eta_{\mathrm{eff},i}=M_i\cdot\eta_{\mathrm{p},0}$ with respect to the reference scan probe efficiency and the measured enhancements $M_i$.
\item[$\#$] Ratio of the observable volumes of interest $V_i$ based on a constant slice thickness and on the various IDs with respect to the reference scan without any lens.
\end{tablenotes}
\end{threeparttable}
\end{adjustwidth}
\end{table}

\FloatBarrier
\subsection{Comparison between model and experiment}

\begin{figure}[h!tbp]
\centering
\includegraphics[]{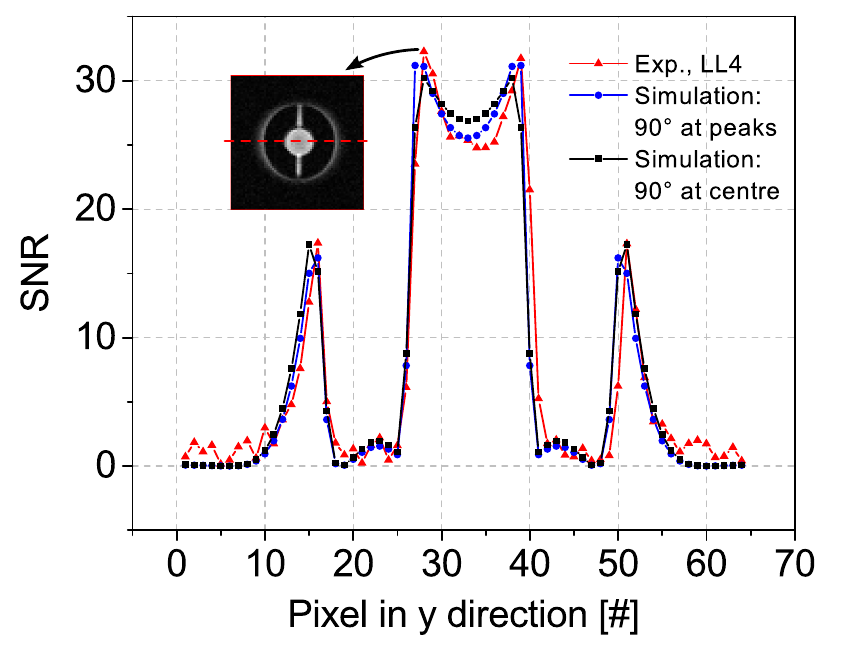}
    \caption{Comparison of the measured SNR profile for LL4 and the simulated SNR profiles for two differently defined \SI{90}{\degree}-pulses: at the highest signal intensity in accordance with the experiment; at the centre-point of the arrangement.}
  \label{fig:FA}
\end{figure}

\begin{figure}[h!tbp]
\centering
\includegraphics[]{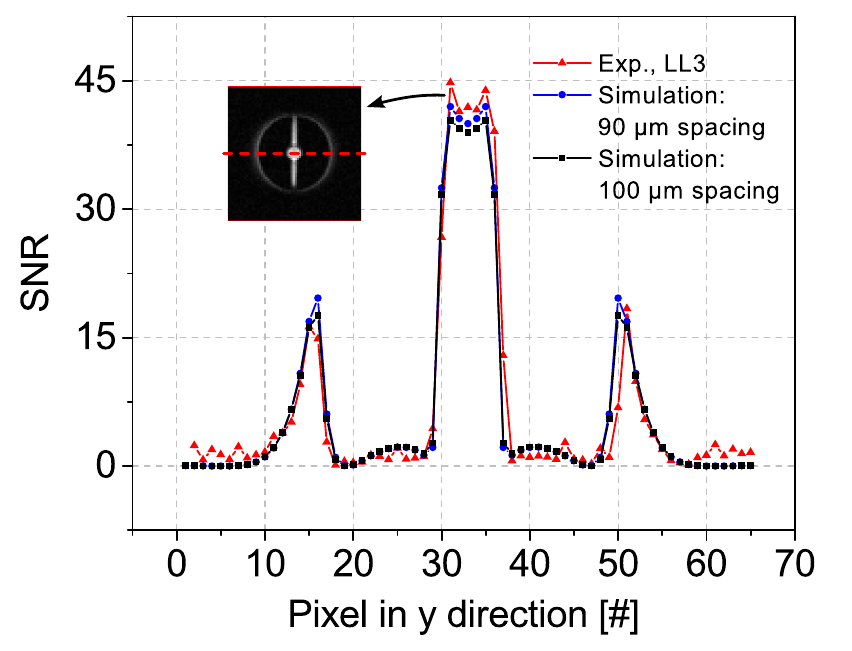}
    \caption{Measured and simulated SNR profiles for LL3 at the \SI{90}{\degree}-pulse power. Simulated profiles were calculated for two different lens spacings (\SIlist{90;100}{\micro\metre}), while in both cases, the \SI{90}{\degree}-pulse was defined at the centre-point.}
  \label{fig:spacing}
\end{figure}

Figs.~\ref{fig:FA} and~\ref{fig:spacing} present comparisons between the experimental results and the corresponding simulations according to equations~(\ref{ILm})-(\ref{ILme}). The simulated SNR profiles were generated by first calculating the field amplification $B_{zA}$ over an array of \num{31 x 65 x 101} points: 31 points across the pixel in the $x$-direction, 65 points along the $y$-axis (i.e., one for each pixel), and 101 points across the slice in the $z$-direction. This field was averaged over the $x$- and $z$-directions, and the value ($B_{zA90}$) either at the highest signal peak or at the centre-point in $y$ was used to define the \SI{90}{\degree} flip angle. A corresponding SNR array of \num{31 x 65 x 101} points was then generated by applying $\mathrm{SNR}=\left|(B_{zA}\sin(\pi/2\cdot B_{zA}/B_{zA90}))\right|$. This SNR array was subsequently averaged over the $x$- and $z$-directions to give the SNR amplification factor along the $y$-axis. Finally, the result was multiplied by the maximum SNR of the reference scan, listed in the first row of Table~\ref{tab:table2}.

Fig.~\ref{fig:FA} displays SNR profiles for LL4 and demonstrates the impact of defining the global \SI{90}{\degree}-pulse either at the location of the largest field amplification (as per the experiment) or at the centre-point (cf. Fig.~\ref{fig:LL4}). Fig.~\ref{fig:spacing} displays SNR profiles for LL3 and demonstrates the impact of error in the spacing of the lenses (i.e. \SI{100}{\micro\meter} versus \SI{90}{\micro\meter}). For both of the simulated profiles in Fig.~\ref{fig:spacing}, the \SI{90}{\degree}-pulse was defined at the centre-point, and for the case with \SI{90}{\micro\meter} spacing the array size used in the calculation was reduced to \num{31 x 65 x 91} points.

\FloatBarrier
\section{Discussion}\label{sec:disc}
Fig.~\ref{simfigs} demonstrates that the magnitude and homogeneity of the focused field is strongly dependent on the number and geometrical parameters of the Lenz lenses. For a single lens, a smaller inner loop results in a stronger field at the cost of homogeneity. However, with the introduction of a second lens this homogeneity can be partially recovered, provided that an appropriate axial separation is chosen. For example, those pairs of lenses in Fig.~\ref{simfigs} for which the ratio of the inner loop radius to the axial separation is equal to one (i.e. akin to the Helmholtz pair; column 1, row 3 and column 2, row 4) clearly display a superior trade-off of amplification to homogeneity, over a spherical ROI, compared to the other arrangements.

However, to obtain the best trade-off over an arbitrarily shaped ROI, especially when considering additional lenses, optimisation via equation~(\ref{opt}) is necessary. Fig.~\ref{optfigs} demonstrates that for an average relative field error of 1\% within the ROI defined in Section~\ref{ssec:meth_simopt}, it is possible to achieve amplifications of 1.5, 2.2 and 3.0 with one, two and four lenses, respectively, over the use of a Helmholtz pair alone. If the field constraint is relaxed to 2\% error, the amplification increases to 1.7, 2.5 and 3.2, respectively (data not shown). Notice that for the case of four lenses in Fig.~\ref{optfigs}, the inner loops lie approximately on the surface of a sphere. However, for the equivalent case with 2\% error, the two pairs of lenses are brought into close proximity to one another (not shown).

Equations~(\ref{ILm})-(\ref{opt}) therefore allow the study and design of a wide variety of lens arrangements with flexible field constraints. Indeed they may be applied to circular lenses of any orientation without necessarily being restricted to the centred coaxial arrangements considered herein. Furthermore, the choice of considering a Helmholtz pair as the transmit/receive RF coil has been made for demonstrative purposes only, since it was also used in some of the experiments, and the theory is applicable to any system for which the mutual inductance can be calculated. Note that we have ignored capacitive effects that may be present for multiple closely-spaced lenses and also the finite gap between the elements that join the inner and outer loops. Both of these factors will reduce the achievable amplification by some degree. Note also that a direct method was used to solve equation~(\ref{opt}), which was somewhat sensitive to the initial guess and hence the two-stage procedure described in Section~\ref{ssec:meth_simopt} was implemented. An alternative approach would be to use stochastic optimisation, such as simulated annealing, to guarantee convergence to the global optimum at the cost of runtime.

The experiments performed in this study clearly confirm the focusing effect and the associated local SNR enhancement as predicted by the model. For the designs fabricated in this study, we achieved enhancements $1.6 \leq M_i \leq 2.8$ (Table~\ref{tab:table2}). For the largest enhancement $M_1$, the power required to generate a \SI{90}{\degree}-pulse was reduced by a factor of \num{5.6} compared to the reference. Although in this case the observable volume of interest was reduced to \num{1/36} of the initial volume, the gain in SNR corresponds to a $2.8^2\approx 7.8$-fold decrease in acquisition time, since $\mathrm{SNR}\propto \sqrt{\mathrm{TA}}$ \cite{macovski_noise_1998}. Higher SNRs were achieved for the plate-based designs due to the lower electrical resistance, i.e., the increased noise contribution in the case of the wire-based design.

Mean SNR enhancement may be estimated using the derived probe efficiencies $\eta_{\mathrm{p},i}$, which are based on the \SI{90}{\degree}-pulse power $P_i$, and these were found to correspond well with the measured effective probe efficiencies $\eta_{\mathrm{eff},i}$ (Table~\ref{tab:table2}).

A trade-off between the enhancement and the homogeneity of the SNR profile was clearly apparent in the experimental results. For example, for LL4 at $\alpha_4=\SI{90}{\degree}$ we observed an enhancement of $M_4=\num{1.6}$ and a standard deviation (STD) of \num{2.8}, whereas at $\alpha_5=\SI{106}{\degree}$ the STD improved by \SI{57}{\percent} to \num{1.2} at the cost of a \SI{14}{\percent} decrease in the enhancement to $M_5=\num{1.4}$ (see also Fig.~\ref{fig:LL4}). In the latter case, $\alpha$ approached \SI{90}{\degree} at the centre-point and exceeded \SI{90}{\degree} at the peak field positions, which occurred close to the inner edge of the Lenz lens defined by the inner diameter, and this resulted in a reduced SNR at these points.

The observed behaviour in the experiments is matched well by the theoretical model, as illustrated in Fig.~\ref{fig:FA} for LL4. For the case in which the \SI{90}{\degree}-pulse corresponds to the peak field positions, the simulations are in high agreement with the results obtained in the experiment. Furthermore, the simulated profile flattens out for the case in which the \SI{90}{\degree}-pulse is at the centre-point, as expected from the experiments, and STD decreases from \numrange{2.3}{1.3}.

Good agreement between simulations and experiment was also found for LL3, as shown in Fig.~\ref{fig:spacing}. The slightly higher SNR values obtained in the experiment may be a result of fabrication tolerances, such as uncertainty in the spacing between lenses. In the present case, such tolerances may occur if the final thickness of the plated lenses is either not uniform or inaccurate, or if the thickness of the photoresist defining the spacing between the two lenses deviates from its nominal value. The example given in Fig.~\ref{fig:spacing} shows that reducing the spacing from \SIrange{100}{90}{\micro\meter} leads to a \SI{3}{\percent} increase in mean SNR. Furthermore, the simulations assume line currents for the field calculation, whereas LL3 and LL4 are constructed using conductive material with a rectangular cross-section that has a low aspect ratio. Therefore, the simulations may represent a lower bound to the trend observed in Fig.~\ref{fig:90deg} from plate-based to wire-based lenses.

To conclude, we have presented a novel method to locally increase the signal-to-noise ratio in nuclear magnetic resonance by using Lenz lenses to focus the RF magnetic field of an NMR coil into a smaller volume of interest. As a result, higher spatial resolutions or reduced acquisition times become possible. The achievable enhancement strongly depends on the size ratios of the excitation coil and the dimensioning of the lens. The use of Lenz lenses enabled not only the magnetic flux density to be intensified, but also the resulting field to be shaped by arranging multiple lenses in a pre-defined manner. The paper presented a theoretical description of a set of Lenz lenses, which conformed closely to the experimental data obtained in a series of imaging experiments. The simulations enabled the optimisation of the design with respect to parameters such as signal amplitude and homogeneity. Lenz lenses are useful not only for amplifying the $B_1$ magnetic field in the sample, but also for attenuating it in other regions, for example to avoid exciting regions with jumps in magnetic susceptibility, which would otherwise lead to distortions of the experiment, or to protect electronics from strong RF radiation.

\FloatBarrier
\section{Methods}
\subsection*{Simulations} All simulations were performed using Matlab\textsuperscript{\textregistered} (R2012b, Mathworks\textsuperscript{\textregistered}).
The optimisation in equation~(\ref{opt}) was achieved using the function {\it fmincon} from the Matlab\textsuperscript{\textregistered} Optimisation Toolbox\texttrademark\thinspace (interior-point algorithm) for three cases: one, two and four lenses.
The optimisations took the order of seconds (1 lens, 7 s; 2 lenses, 25 s; 4 lenses, 106 s) on a standard desktop computer (2.7 GHz Intel\textsuperscript{\textregistered} Core\texttrademark\thinspace  i7 processor with 16 GB of RAM).

\subsection*{Device fabrication}
\begin{figure}[h!tb]
\centering
\includegraphics[]{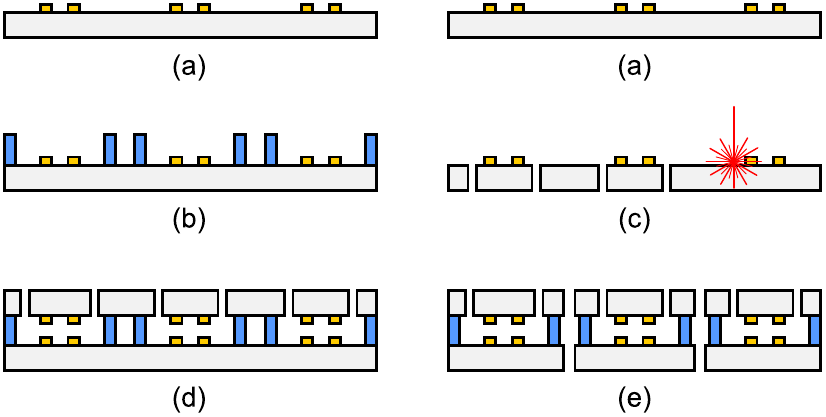}
\caption{Fabrication steps of the double Lenz lens chips. (a) Au-electroplating of lenses on two glass substrates. (b) Lamination and structuring of dry film photoresist on first wafer. (c) UV-laser drilling of microfluidic ports into second wafer. (d) Adhesive full-wafer bonding of both substrates. (e) Dicing of individual chips.}
\label{fig:ll-fab}
\end{figure}

Fig.~\ref{fig:ll-fab} illustrates the fabrication steps.

At first, a \SI{15/150}{\nano\meter} Cr/Au seed layer was evaporated on two \SI[separate-uncertainty]{208\pm{}20}{\micro\meter} thick, 4-inch diameter float glass wafers (D~263\textsuperscript{\textregistered{}}~T, Schott AG, Mainz, Germany), where the Cr-layer served as an adhesion promoter between the glass surface and the Au-layer. Subsequently, Hexamethyldisilazane (HMDS) was applied from the gas phase before spin-coating a \SI{20}{\micro\meter} layer of AZ9260 positive-tone photoresist (MicroChemicals GmbH, Germany). The wafers were stored in ambient atmosphere for around \SI{4}{\hour} for rehydration, before windows for electroplating were opened using UV photo-lithography, which served as a mould structure. In the subsequent electroplating step, alignment marks and Lenz lenses were structured accordingly by depositing a \SI{5}{\micro\meter} thick Au-layer. In the case of the wire-type lenses, the width of the conductor was \SI{50}{\micro\meter}. An SEM-closeup of an electroplated, wire-type Lenz lens is presented in Supplementary Fig.~2.

Afterwards, the mould layers were stripped using acetone before the Au seed layer was etched using a potassium iodine/iodide based etchant. All etching and stripping steps were performed using megasonic agitation to ensure uniform wetting of the surface and homogeneous etching rates.

Before the subsequent UV-laser drilling step, dicing foil was laminated on both sides of one wafer to protect the substrate from debris during laser ablation. The latter was done using a UV-laser (TruMark 6330, Trumpf, Germany), where 108 holes were drilled to realise two microfluidic ports for each of the 54 chips. After drilling, the fragile wafer was placed in a petri dish with isopropyl alcohol (IPA), which dissolved the adhesive of the dicing foil, before the wafer was rinsed in deionized (DI) water and dried using nitrogen gas.

Two layers of Ordyl SY355 dry film resist (Elga Europe s.r.l., Milano, Italy) were laminated onto the second substrate using a hot roll laminator (Mylam 12, GMP, Polch, Germany). Lamination was done at a speed of \SI{1}{\centi\meter\per\second}, a pressure of \SI{1}{\bar}, and a temperature of approximately \SI{100}{\celsius}, resulting in a total nominal resist-thickness of \SI{110}{\micro\meter}. The resist was UV exposed at \SI{180}{\milli\joule\per\square\centi\meter} using a mask aligner (MA6, Karl Suss, Germany) to pattern microfluidic channels. After exposure, a post-exposure bake (PEB) was performed for \SI{1}{\minute} at \SI{85}{\celsius}. The structures were developed for about \SI{6}{\minute} in BMR developer using ultrasonic agitation, followed by rinsing in IPA and DI water. Finally, the wafer was spin-dried.

Both substrates were subsequently bonded in a full wafer bonding process \cite{vulto_full-wafer_2009,muller_optofluidic_2010} using a substrate bonder (SB6, Karl Süss, Germany) to realise the double Lenz lens configuration, i.e., parallel pairs of lenses. The wafers were aligned manually and fixed using the clamps of the SB6 chuck before loading it into the machine, where a tool pressure of \SI{2.4}{\bar} was applied for \SI{30}{\minute} at \SI{95}{\celsius}. The stack was hard baked for \SI{2}{\hour} at \SI{150}{\celsius} before dicing \num{54} chips with a nominal size of \SI[product-units = brackets-power]{5 x 16 x 0.5}{\milli\meter} ($\mathrm{width}\times\mathrm{length}\times\mathrm{height}$). The resulting spacing between the lenses of \SI{\approx 100}{\micro\meter} was hence defined by the thickness of the deposited Au-layer and the height of the previously patterned Ordyl photoresist layers.

\subsection*{Magnetic resonance imaging}
All magnetic resonance imaging experiments were performed on an Avance III NMR system (Bruker, Rheinstetten, Germany), controlled by the ParaVision\textsuperscript{\textregistered{}} 5.1 imaging software (Bruker). The NMR scanner was operated at the proton Larmor frequency of \SI{500.13}{\mega\hertz} in combination with a Micro 5 micro-imaging probe base and a Micro 5 gradient system, driven at \SI{40}{\ampere}, which results in a maximum gradient strength of \SI{2}{\tesla\per\metre}.

Throughout the experiments, the attenuation (ATT) was varied from \SIrange{70}{50}{\dB} ($\SI{0.025}{\watt}\leq{}P\leq{}\SI{2.5}{\watt}$) in steps of \SI{0.5}{\dB}. The relationship between ATT and $P$ is given by 
\begin{equation}
P=10^{\frac{\mathrm{ATT}_0-\mathrm{ATT}}{10\,\mathrm{dB}}}P_0,
\end{equation}
where $P_0=\SI{1}{\watt}$ and $\mathrm{ATT}_0=\SI{54}{\dB}$. Acquisition parameters were set to: repetition time $\mathrm{TR}=\SI{500}{\milli\second}$, echo time $\mathrm{TE}=\SI{5.3}{\milli\second}$, flip angle $\alpha=\SI{90}{\degree}$, effective slice thickness $\mathrm{SI}=\SI{100}{\micro\meter}$, field of view $\mathrm{FOV}=(\SI{1.92}{\mm})^2$, matrix $\mathrm{MTX}=\num{64 x 64}$ and hence \SI[product-units = brackets-power]{30 x 30}{\square\micro\meter} in-plane resolution, number of averages $\mathrm{NEX}=\num{4}$ and an acquisition time (TA) per scan of \SI{2}{\minute} \SI{8}{\second}.


\begin{thebibliography}{10}

\bibitem{webb_radiofrequency_1997}
A.~G. Webb.
\newblock Radiofrequency microcoils in magnetic resonance.
\newblock {\em Progress in Nuclear Magnetic Resonance Spectroscopy},
  31(1):1--42, July 1997.

\bibitem{lacey_high-resolution_1999}
M.~E. Lacey, R.~Subramanian, D.~L. Olson, A.~G. Webb, and J.~V. Sweedler.
\newblock High-resolution {NMR} spectroscopy of sample volumes from 1 {nL} to
  10 {\textmu{}l}.
\newblock {\em Chemical Reviews}, 99(10):3133--3152, October 1999.

\bibitem{hoult_signal--noise_1976}
D.~I. Hoult and R.~E. Richards.
\newblock The signal-to-noise ratio of the nuclear magnetic resonance
  experiment.
\newblock {\em Journal of Magnetic Resonance}, 24(1):71--85, October 1976.

\bibitem{peck_design_1995}
T.~L. Peck, R.~L. Magin, and P.~C. Lauterbur.
\newblock Design and analysis of microcoils for {NMR} microscopy.
\newblock {\em Journal of Magnetic Resonance, Series B}, 108(2):114--124,
  August 1995.

\bibitem{hill_limits_1968}
H.~D.~W. Hill and R.~E. Richards.
\newblock Limits of measurement in magnetic resonance.
\newblock {\em Journal of Physics E: Scientific Instruments}, 1(10):977,
  October 1968.

\bibitem{olson_high-resolution_1995}
D.~L. Olson, T.~L. Peck, A.~G. Webb, R.~L. Magin, and J.~V. Sweedler.
\newblock High-resolution microcoil {1H-NMR} for mass-limited, nanoliter-volume
  samples.
\newblock {\em Science}, 270(5244):1967--1970, December 1995.

\bibitem{schoenmaker_magnetic_2013}
J.~Schoenmaker, K.~R. Pirota, and J.~C. Teixeira.
\newblock Magnetic flux amplification by lenz lenses.
\newblock {\em Review of Scientific Instruments}, 84(8):085120, August 2013.

\bibitem{HOULT:2000fd}
D.~I. Hoult.
\newblock {The Principle of Reciprocity in Signal Strength Calculations -- A
  Mathematical Guide}.
\newblock {\em Concepts in Magnetic Resonance}, 12(4):173--187, 2000.

\bibitem{HOULT:2011jv}
D.~I. Hoult.
\newblock {The Principle of Reciprocity}.
\newblock {\em Journal of Magnetic Resonance}, 213(2):344--346, December 2011.

\bibitem{RamoWV1994}
S.~Ramo, J.~R. Whinnery, and T.~Van~Duzer.
\newblock {\em Fields and Waves in Communication Electronics}.
\newblock John Wiley \& Sons, Inc., 3rd edition, 1994.

\bibitem{Wheeler1942PIRE}
H.~A. Wheeler.
\newblock Formulas for the skin effect.
\newblock {\em Proceedings of the IRE}, 30(9):412--424, 1942.

\bibitem{Knight2013}
D.~W. Knight.
\newblock Practical continuous functions for the internal impedance of solid
  cylindrical conductors.
\newblock http://www.g3ynh.info/zdocs/comps/Zint.pdf (version 2.06.), Apr 2013.

\bibitem{Dengler2013}
R.~Dengler.
\newblock Self inductance of a wire loop as a curve integral.
\newblock arXiv:1204.1486v2 [physics.class-ph], Sep 2013.

\bibitem{BabicSAG2010IEEETM}
S.~Babic, F.~Sirois, C.~Akyel, and C.~Girardi.
\newblock Mutual inductance calculation between circular filaments arbitrarily
  positioned in space: alternative to grover's formula.
\newblock {\em IEEE Transactions in Magnetics}, 46(9):3591--3600, 2010.

\bibitem{spengler_heteronuclear_2016}
N.~Spengler, J.~Höfflin, A.~Moazenzadeh, D.~Mager, N.~MacKinnon, V.~Badilita,
  U.~Wallrabe, and J.~G. Korvink.
\newblock Heteronuclear micro-helmholtz coil facilitates µm-range spatial and
  sub-hz spectral resolution nmr of nl-volume samples on customisable
  microfluidic chips.
\newblock {\em PLoS ONE}, 11(1):e0146384, 01 2016.

\bibitem{macovski_noise_1998}
A.~Macovski.
\newblock Noise in mri.
\newblock {\em Magnetic Resonance in Medicine}, 36(3):494--497, 1996.

\bibitem{vulto_full-wafer_2009}
P.~Vulto, T.~Huesgen, B.~Albrecht, and G.~A. Urban.
\newblock A full-wafer fabrication process for glass microfluidic chips with
  integrated electroplated electrodes by direct bonding of dry film resist.
\newblock {\em Journal of Micromechanics and Microengineering}, 19(7):077001,
  July 2009.

\bibitem{muller_optofluidic_2010}
P.~Müller, N.~Spengler, H.~Zappe, and W.~Mönch.
\newblock An optofluidic concept for a tunable micro-iris.
\newblock {\em Journal of Microelectromechanical Systems}, 19(6):1477--1484,
  December 2010.

\end{thebibliography}

\section*{Acknowledgements}
This work was partially funded by the European Research Council (ERC) grant no. 290586 NMCEL (N.S., M.V.M., and J.G.K.) and by an Alexander von Humboldt research fellowship (P.T.W.). We further acknowledge Michael Pauls for recording the SEM images, Kay Steffen for electroplating, and Volker Lehmann from Bruker BioSpin for advice with the NMR instrumentation.

\FloatBarrier
\section*{Author contributions}
N.S. and J.G.K. developed the concept of the study. N.S., M.V.M., and U.W. planned the microfabrication. N.S. and M.V.M. conducted the microfabrication, and performed the NMR experiments. P.T.W. derived the theory and performed the numerical optimisation experiments. All authors performed analysis and data interpretation, and drafted the manuscript.

\FloatBarrier
\section*{Competing financial interests}
The authors declare no competing financial interests.

\newpage

\specialsection*{Supplementary Information}

\renewcommand{\figurename}{Supplementary Figure}
\setcounter{figure}{0}

\begin{figure}[h!tbp]
\centering
\textbf{Overview of the experimental setup and summary of the most significant MR images acquired}\par\medskip
\includegraphics[width=\textwidth]{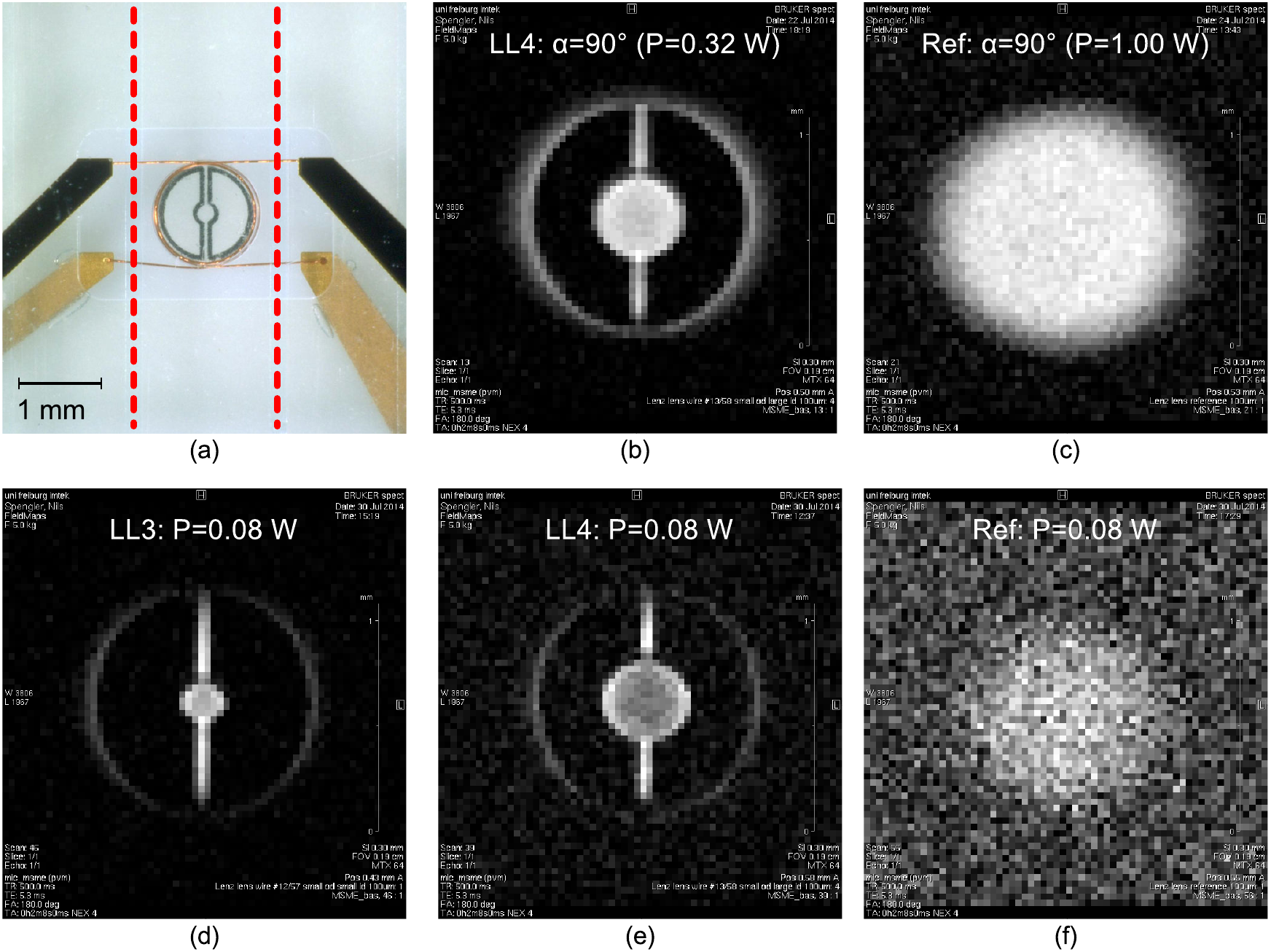}
    \caption{The figure presents the experimental setup and summarises the most significant results obtained. (a) Photograph of a LL3 chip inserted in between the \SI{1.2}{\mm} diameter micro Helmholtz coil pair used throughout the experiments. The microfluidic chamber filled with DI-water is depicted by broken red lines. (b) Acquired MR image using a LL4 chip with a \SI{90}{\degree} pulse at \SI{0.32}{\watt}. (c) \SI{90}{\degree} reference scan without Lenz lens chip, which required a threefold higher power of \SI{1.00}{\watt}. (d) to (f): MR images acquired at a constant power of \SI{0.08}{\watt} to demonstrate the increased SNR for LL3 (d) and LL4 (e) compared to the reference scan without Lenz lens (f).}
  \label{fig:ll-comp}
\end{figure}

\begin{figure}[h!tbp]
\centering
\textbf{SEM closeup of an electroplated LL4 lens}\par\medskip
\includegraphics[]{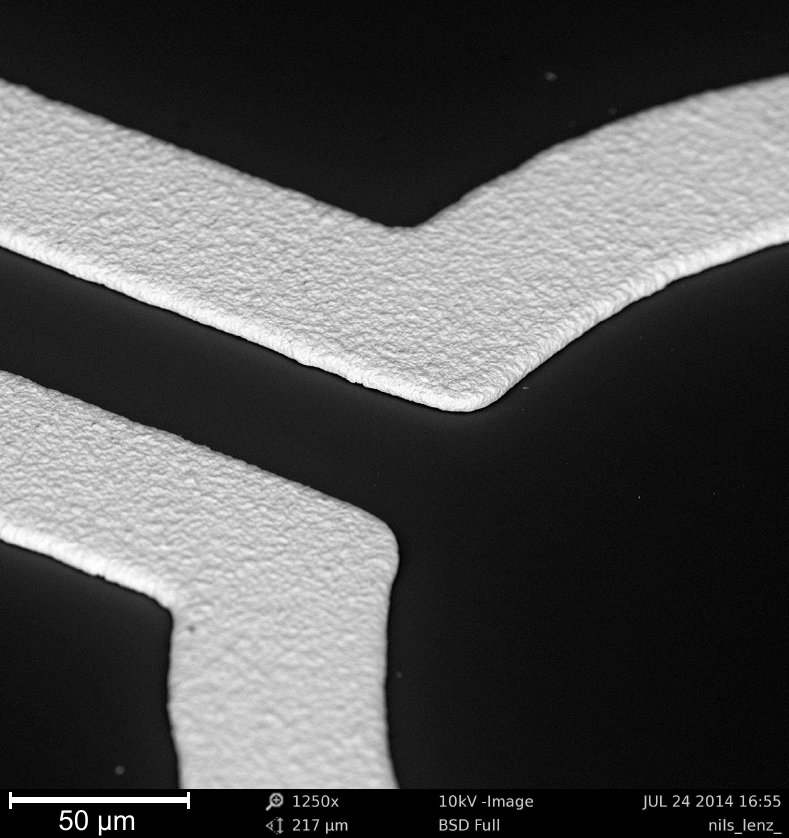}
    \caption{The figure presents an SEM closeup of an electroplated LL4 lens.}
  \label{fig:ll-sem}
\end{figure}

\end{document}